\documentclass[12pt,preprint]{aastex}
\shorttitle{Mid-IR Imaging in $\rho$ Oph}
\shortauthors{Barsony, Ressler, \& Marsh}
\begin{document}
\title{A Mid-Infrared Imaging Survey of Embedded Young Stellar Objects in the $\rho$ Ophiuchi Cloud Core}
%
\author{Mary Barsony\altaffilmark{1,2,3} }
\affil{Department of Physics \& Astronomy, San Francisco State University\\
1600 Holloway Drive, San Francisco, CA  94132}
\email{mbarsony@stars.sfsu.edu}

\author{Michael E. Ressler\altaffilmark{2,3}}
\affil{Jet Propulsion Laboratory, Mail Stop 169-327 \\ 
4800 Oak Grove Drive, Pasadena, CA  91109}
\email{Michael.E.Ressler@jpl.nasa.gov}

\and

\author{Kenneth A. Marsh}
\affil{Jet Propulsion Laboratory, Mail Stop 168-414\\
4800 Oak Grove Drive, Pasadena, CA  91109}
 \email{Kenneth.A.Marsh@jpl.nasa.gov}

\altaffiltext{1}{and Space Science Institute, 
4750 Walnut Street, Suite 205, Boulder, CO  80301}

\altaffiltext{2}{Visiting Astronomer at the W. M. Keck Observatory, 
which is operated as a scientific partnership among the California Institute of Technology, 
the University of California, and the National Aeronautics and Space Administration. 
The Observatory was made possible by the generous financial support of the W. M. Keck Foundation.}

\altaffiltext{3}{Observations with the Palomar 5 m telescope were obtained under 
a collaborative agreement between Palomar Observatory and the Jet Propulsion Laboratory.}
\begin{abstract}
Results of a  comprehensive, new, ground-based mid-infrared imaging survey of the young stellar population of the $\rho$ Ophiuchi cloud are presented.  Data were acquired at the Palomar 5-m and at the Keck 10-m  telescopes with the MIRLIN and LWS instruments, at 0.5$^{\prime\prime}$ and 0.25$^{\prime\prime}$ resolutions, respectively.  
Of 172 survey objects, 85 were detected. Among the 22 multiple systems observed,
15 were resolved and their  individual component fluxes determined.
A plot of the frequency distribution
of the detected objects with SED spectral slope shows that YSOs spend $\sim$ 4 $\times$ 10$^5$
yr in the Flat Spectrum phase, clearing out their remnant infall envelopes. Mid-infrared variability
is found among a significant fraction of the surveyed objects and is found to occur
for all SED classes with optically thick disks. 
Large amplitude near-infrared variability, also found for all SED classes with optically
thick disks, seems to occur with somewhat higher frequency at the earlier evolutionary stages.
Although a general trend of mid-infrared excess and NIR veiling exists progressing
through SED classes, with  Class I  objects generally exhibiting $r_K \ge$ 1, Flat Spectrum objects with
$r_K \ge 0.58$, and Class III objects with $r_K=0$, Class II objects exhibit the 
widest range of $r_K$ values, ranging from $0 \le r_K \le 4.5$.  However, the 
highly variable value of veiling
that a single source can exhibit in any of the SED classes in which active disk
accretion can take place is striking, and 
is direct observational evidence for highly time-variable accretion activity in disks.
Finally, by comparing mid-infrared vs. near-infrared excesses in a subsample with
well-determined effective temperatures and extinction values, disk clearing mechanisms
are explored.  The results are consistent with disk clearing proceeding from the inside-out.

\end{abstract}


\section{Introduction}

The $\rho$ Ophiuchi cloud core continues to be the subject of intense study at all wavelengths,
since it harbors the nearest  site of the formation of several
hundred young stellar objects (YSOs).   Mid-infrared observations of $\sim$ 2 dozen YSOs 
in the $\rho$ Oph cloud core resulted in the development of the currently accepted classification
scheme for such objects \cite{lad84,lad87}.
The most comprehensive previous ground-based mid-infrared survey of  the $\rho$ Oph
clouds included  56 objects, 49 of which were in the central L1688 cloud core \cite{gre94}.
A survey of the entire YSO population of the $\rho$ Oph cloud core became practicable
for the first time with the advent of  mid-infrared arrays mounted on large-aperture ground-based telescopes towards the end of  the last decade. It is just such a survey that is the subject
of this work.

Mid-infrared studies of YSOs are especially relevant for studies 
of young, planet-forming disks, since mid-IR emission peaks at distances of order $\sim$ 1 AU
from the central object. By contrast, near-infrared continuum (NIR) emission from hot dust originates
closer to the central YSO  \cite{eis03,mst01,mst99}. Comparison of the near- and mid-infrared
properties of young disks is useful for understanding disk structures \cite{chi97}, whereas
multi-epoch comparison of mid-infrared fluxes with previous \cite{bon01} and future (Spitzer Legacy
and GTO) observations will aid our understanding of disk accretion processes.
 
Of obvious importance for planet formation are not only studies of average YSO disk lifetimes
e.g., \cite{hai01}, but also studies of disk dissipation processes \cite{arm03}.  Comparative studies
of the population of very young ($\le$ 1 MYr), diskless YSOs \cite{wil01} with the equally
young population of YSOs with optically thick disks detected in the present survey, should greatly
aid our understanding of the reason for the observed widely varying disk dissipation 
timescales.

Finally, no current space-based observatory can match the angular resolution
achievable from the ground at mid-infrared wavelengths, which is ideal for 
studying young binaries and their disks at the distance to the $\rho$ Oph clouds.
Since the majority of young stars form in binary/multiple systems, studies of planet-forming
disks must, of necessity, include studies of the disks in multiple systems.
The binary separation of YSOs at the distance
to the $\rho$ Oph clouds peaks near a projected angular separation of  0.25$^{\prime\prime}$
\cite{bar03}, which is the diffraction-limit of the Keck 10-meter telescopes at 10 $\mu$m.

In the following, we describe the source selection criteria for our mid-infrared survey, 
along with the data acquisition and data reduction procedures
in $\S$ 2.   The key results of our new, mid-infrared survey are presented in $\S$ 3 and the
implications of these are discussed in $\S$ 4.

\section{Source Selection Criteria, Observations, and Data Reduction}

The starting point for our target list was the 
near-infrared imaging survey \cite{bkl97} of the central square degree
of the $\rho$ Ophiuchi cloud core (also known as L1688), which catalogs 4495 objects.
To winnow this source list to a manageable number of target objects,
while  maximizing  the likelihoods
of  cloud membership and of mid-IR detectability, we imposed
combined near-infrared brightness
and color criteria.  Target objects for this study were, therefore, required to be bright at 2.2 $\mu$m
(K$\le$13.0), since YSOs are generally bright NIR emitters, even when highly reddened, and had
to have very red NIR color (H$-$K $\ge$ 1.67), since high near-infrared excess is
generally a disk  indicator.

Since $\rho$ Oph is, for the most part, a low-mass star-forming cloud,
at its distance of 140 pc, the apparent magnitude range $5 \le K \le 10.5$
essentially encompasses its entire population, not allowing for
extinction. According to theoretical pre-main-sequence isochrones, 
one million year old YSOs at the hydrogen-burning limit
of $\approx$ 0.08 M$_{\odot}$ should have an apparent $K=10.5$
at this distance, with no foreground extinction \cite{wgm99}. At the other end of the mass spectrum, a similarly youthful Herbig Ae star of 
4 M$_{\odot}$, WL16, is observed to have an apparent $K=7.92$ in $\rho$ Oph,
seen through $A_V=31$ \cite{res03}.

Using the relation $A_K=1.4(H-K)$, our color selection
criterion of $H-K\ge1.67$, would result in $A_K\ge2.3$, assuming
an intrinsic $(H-K)_0 = 0.0$ (Cohen et al. 1981).
Using the relation, $A_V = 15.4 \times [(H-K)_{obs} - (H-K)_0]$,
and noting that $(H-K)_0 = 0.3$ for an M dwarf with no infrared excess,
our color selection criterion of $H-K\ge 1.67$, when combined with our 
apparent brightness criterion of $K \le 13$,
guarantees that we are sampling the cloud core's embedded
population through $A_V \sim 21$ even for young brown dwarfs.
At the other end of the brightness scale, these selection criteria
could also include occasional stray background K and M giants out to $\sim$ 1 kpc,
depending on the actual obscuration along the line-of-sight.  
In fact, the most recent NIR spectroscopic survey of the $\rho$ Ophiuchi YSO
population has shown that of the 80
objects in common to both surveys, only six of our target
list (Elias 35, GY232, VSSG6, GY65, GY45, BKLT J162618$-$242818) are confirmed
background K or M giants \cite{luh99}.

The combined color and brightness criteria described above, when applied to the BKLT survey,
selected 104 of 4495 objects, 
of which all but 7  (GY12, GY30, IRS 26, BKLT 162805$-$243354, GY 312,
BKLT 162629$-$241748, and BKLT 162735$-$244628) were observed by us
in the mid-infrared. This source list was supplemented with objects that were 
known or suspected to be cloud members from various optical emission-line,
near-infrared, and far-infrared surveys, but which did not satisfy the 
above-mentioned combined color and brightness criteria 
\cite{str49,dol59,vss75,eli78,wil83,ylw86,wly89,leo91,gyo92,bkl97,crb93}.
In total, we report mid-infrared fluxes or flux upper limits for 172 individual objects.  
The spatial distribution of the objects targeted by this survey is plotted in Figure 1, overlaid
on the contours of  C$^{18}$O (J$=$2$\rightarrow$1) emission from the cloud core, for reference
\cite{wil83}.

The mid-infrared imaging data for this survey of the $\rho$ Oph
cloud core's embedded population were obtained with MIRLIN, 
JPL's 128 $\times$ 128 pixel Si:As camera
\cite{res94}, and the University of California's
Long Wavelength Spectrometer (LWS) \cite{jop93}. 
The relevant observing and data reduction parameters 
are listed in Table 1. The first column lists
the UT date of observation, the second column specifies
the telescope used.  All observations at Palomar and on Keck II
were made with MIRLIN, whereas LWS was the instrument used on Keck I.

The MIRLIN data were acquired at N-band 
($\lambda_0=10.78\ \mu$m, $\Delta \lambda=5.7\ \mu$m). 
However, in order to avoid saturating the 
medium-well-depth Si:As detector of the LWS instrument,
a narrower band, 12.5 $\mu$m filter 
($\lambda_0=12.5\ \mu$m, $\Delta \lambda=1.0\ \mu$m) 
was used.
MIRLIN has a plate-scale of 0.138$^{\prime\prime}$/pixel
and a 17.7$^{\prime\prime}$ $\times$ 17.7$^{\prime\prime}$
field of view at the Keck II telescope;
corresponding values for MIRLIN at the Palomar 5-meter
are 0.15$^{\prime\prime}$/pixel and 
19.2$^{\prime\prime}$ $\times$ 19.2$^{\prime\prime}$, respectively.
At the Keck I telescope, the LWS instrument has a plate-scale of 
0.08 $^{\prime\prime}$/pixel and a 
10.2$^{\prime\prime}$ $\times$ 10.2$^{\prime\prime}$
field-of-view. 
For reference, the full-width at half-maximum of a diffraction-limited
image at $N$ band is $\sim$ 0.25$^{\prime\prime}$ at the Keck Telescopes 
and 0.47$^{\prime\prime}$ at the Palomar 5-meter.

Data were acquired with traditional mid-IR chopping and
nodding techniques. Specifically, for MIRLIN observations, 
the telescope's secondary mirror was chopped 8$^{\prime\prime}$ 
in a north-south direction,
at a rate of a few Hz; then the entire telescope was nodded 
8$^{\prime\prime}$ east-west, in order to remove residual differences 
in the background level. Total on-source integration times were 
typically 24-25 seconds at each telescope for the program sources.
On-source integration times consisted of several hundred to a thousand
coadded chop pairs, with 5-6 msec integration times per frame. 
For the LWS observations, the secondary mirror chop throw was 4$^{\prime\prime}$,
as was the telescope nod, 4$^{\prime\prime}$
alternately on either side of the source along a straight line. 
The LWS chop frequency was 4.8 Hz, integration times per frame were 15 msec, 
and typical total on-source integration times were 72 seconds.
All raw images were background-subtracted,
shifted, and co-added with our in-house IDL routine, ``mac'' (match-and-combine).

The flux standards observed for each night are listed in
Column 3 of Table 1, along with corresponding N band magnitudes
in Column 4. These were also used for airmass monitoring.
Photometry for the standard stars was performed in circular software
apertures with radii (in arcsecond units) indicated in Column 5 of Table 1.
A straight-line fit to the instrumental minus true magnitudes of the standards
as a function of airmass for each night resulted in the determination of the 
airmass corrections and zero-point offsets listed in Columns 8 and
9 of Table 1, respectively. 
No airmass corrections were used for the Keck data, whereas typical 
airmass corrections at Palomar were of order 0.1$-$0.2 mags/airmass. 
Photometric consistency between all standards during a given night's 
observing was typically of order $\pm$ 0.05 magnitudes. By adding the 
errors in the zero-point offsets, the airmass
corrections, the aperture corrections, and the uncertainties in the magnitudes
of the standards in quadrature and taking the square root, we estimate the
total photometric accuracy of the Keck data to be good to $\pm$ 0.06 magnitudes,
and the Palomar data to $\pm$ 0.08 magnitudes.  To convert these errors
to Janskys, note that 0.00 magnitudes at $N$-band with MIRLIN corresponds
to 33.4 Jy, whereas 0.00 magnitudes at 12.5$\mu$m with LWS corresponds to 
25.2 Jy. 

Photometry for program sources was typically performed in
the software apertures listed in Column 6 of Table 1, under the column
heading, ``Target Object Aperture Radius.'' Aperture corrections were 
derived from the flux standards for each night, and applied to each target 
object's instrumental magnitude before application of the 
zero-point calibration and airmass correction. The aperture
corrections for each night's observing are listed in 
Column 7 of Table 1.
For bright sources with N $\le$ 3.5,
the same software apertures were used as for the bright standards.
For the case of sources so extended that they appear significantly brighter
in a larger software aperture, a software aperture large enough to include
all of the detected flux was used.

In the case of close companions, the combined system flux
was first determined from a software aperture chosen to be
large enough to contain both objects.
Subsequently,  the relative photometry of each component was determined
by fitting a known point-source calibrator 
(generally one of the flux calibrators observed that night) to the 
individual source peaks.  The total flux was then divided amongst the
components in the ratio determined by the relative point-source fitting
photometry.

Upper limits were calculated in the same apertures as faint target objects.
The resultant background subtracted counts/sec were multiplied by a 
factor of three, and the result converted to Jy units.   Averaging the upper limits
obtained in this manner for each telescope/instrument combination, it is found that the mean 
1 $\sigma$ flux errors were 0.018 $\pm$ 0.024 Jy
at  12.5 $\mu$m with LWS at Keck I, 0.008 $\pm$ 0.005 Jy at 10.8 $\mu$m with MIRLIN at Keck II,
and 0.029 $\pm$ 0.033 Jy at 10.8 $\mu$m with MIRLIN at  Palomar.

\section{Results}

%

The mid-infrared photometry for our target objects is presented in Table 2.
The list is R.A. ordered. The first column of Table 2 lists the BKLT source name,
and the third and fourth columns list the objects' J2000 coordinates from Barsony et al. 1997. 
For reference, the second  column of Table 2 lists a common, alternative alias
for each BKLT  object (for cross-references
of aliases for infrared sources, see Barsony et al. 1997).
The fifth and sixth columns list each object's mid-infrared flux or flux upper limit 
in Jy units (either at 10.8 $\mu$m for MIRLIN data, or at 12.5 $\mu$m for LWS data). 
The seventh column lists the UT date of observation, 
and the eighth and last column lists the telescope/instrument combination used.
In some cases, several entries exist for the same object, if it was observed
on more than one occasion. In such instances, 
the different measurements are separately tabulated.

In Figures 2 \& 3, we compare our results with {\it ISOCAM} photometry 
\cite{bon01}. These authors list photometry for 212 objects,
of which 199 are in the L1688 cloud core, and the rest are in L1689.
There are 120 members of L1688 common to both surveys. 
Figure 2 is a plot of detected objects from Table 2 that are
in common with {\it ISOCAM}-detected objects. 
Figure 3 is a plot of sources with upper limits listed in Table 2
that are in common with {\it ISOCAM}-detected objects. 

In Figure 2, we compare our ground-based mid-infrared photometry
for 69 sources that were detected in both {\it ISOCAM} filters.  The published 
 {\it ISOCAM} fluxes are for filters centered at 6.7 $\mu$m
and 14.3 $\mu$m.  For comparison with MIRLIN data, whose broadband N filter
is centered at 10.8 $\mu$m, we linearly interpolated the published {\it ISOCAM} 
fluxes to 10.8 $\mu$m. These objects are represented by filled diamonds in Figure 2.
For comparison with LWS data, whose narrower filter was centered at 12.5 $\mu$m,
we linearly interpolated the published {\it ISOCAM} 
fluxes to 12.5 $\mu$m. These objects are represented by the filled squares in Figure 2.
The straight line plotted in Figure 2 represents perfect agreement between the measurements 
reported in this survey and the interpolated measurements from {\it ISOCAM}.
The agreement is generally good between the two datasets. This means that 
the ground-based fluxes agree with the {\it ISOCAM} fluxes to within 3$\sigma$
of the published uncertainties. For the sources with good agreement between
the ground-based and {\it ISOCAM} fluxes, we can infer that any extended emission
on scales $\le$ 18$^{\prime\prime}$ is below our detection limit, since
18$^{\prime\prime}$ diameter apertures were used for the {\it ISOCAM} flux measurements
(except for the faintest sources, for which a 9$^{\prime\prime}$ diameter aperture was used),
whereas a much smaller software aperture was used for the ground-based flux measurements
(see Table 1).  Sources whose measured fluxes have larger discrepancies between
the ground-based and {\it ISOCAM} values are either intrinsically variable or have significant
extended emission. These are listed in Table 3 and discussed
in $\S$ 4.2.

Four objects in common to both surveys, BKLT J162609$-$243411 (SR-3) , 
BKLT J162634$-$242330 (S1), BKLT J162702$-$243726 (WL16), and 
BKLT J162659$-$243458 (WL22),
do not have {\it ISOCAM} fluxes available, although they are strong
mid-infrared emitters. All four are inferred to be embedded early-type stars 
associated with extended mid-infrared emission \cite{bon01}. 
For SR-3, a star of spectral type A0 \cite{str49}, we find a 10.8 $\mu$m flux of 0.15 Jy 
(see Table 2), whereas its previously published N-band flux, 
observed through a 6$^{\prime\prime}$ or 8$^{\prime\prime}$ aperture,
was 0.36 Jy \cite{lad84}. These findings are consistent with the presence of extended mid-infrared
emission for this source. For S1, a star of B4 spectral type \cite{bou92}, we find
a 10.8 $\mu$m flux of 0.065 Jy, corresponding to N$=$6.77 (see Table 2). 
Previous ground-based mid-infrared photometry for this object was 4.9 magnitudes
at 11.1 $\mu$m in an unspecified, but presumably larger, aperture \cite{vss75},
consistent with the presence of extended mid-IR emission centered on S1.  This object
is listed as displaying a double-peaked spectral energy distribution 
by Wilking et al. (2001).
A detailed, spatially-resolved study of the embedded Herbig Ae star,
WL16, shows that the extended mid-infrared emission from this source
originates from a $\sim$ 900 AU diameter inclined disk
composed of PAH (polycyclic aromatic hydrocarbon) and VSG (very small grain) particles, 
obscured by a foreground extinction of A$_V=31$ magnitudes \cite{res03}.
WL22 is inferred to be of early spectral type due to its relatively
high infrared luminosity, strong, extended, mid-infrared emission,
and lack of millimeter continuum flux.

Figure 3 is a plot of the objects common to both surveys,
for which only upper limits are available from the ground-based observations reported here. 
Filled-diamond symbols indicate 10.8 $\mu$m flux upper limits derived from MIRLIN observations; 
filled squares indicate 12.5 $\mu$m flux upper limits derived from LWS observations.
Downward pointing arrows represent the {\it ISOCAM} upper limits. 
Again, the published  {\it ISOCAM} fluxes are for filters centered at 6.7 $\mu$m
and 14.3 $\mu$m. 
For cases where both 6.7 $\mu$m and 14.3 $\mu$m 
{\it ISOCAM} fluxes are available, we have linearly interpolated the {\it ISOCAM} fluxes 
to 10.8 $\mu$m for direct comparison with the MIRLIN data (filled diamonds) and 
to 12.5 $\mu$m, for direct comparison with the LWS data (filled squares).
For those cases where only 6.7 $\mu$m {\it ISOCAM} fluxes were available, we have 
taken the quoted 15 mJy completeness limit at 14.3 $\mu$m as if it were
the actually measured 14.3 $\mu$m flux, to derive strict interpolated 
{\it ISOCAM} upper limits at 10.8 $\mu$m and 12.5 $\mu$m for comparison
with the ground-based upper limits.  
In Figure 3, the straight line is a locus of points
representing perfect agreement between ground-based photometry
and the interpolated {\it ISOCAM} fluxes. Only one object, GY 301,
has an inferred 10.8 $\mu$m {\it ISOCAM} flux that differs significantly from
the ground-based measurement at this wavelength.

There remain 78 members of the L1688 cloud core for which
only {\it ISOCAM} photometry is available.  Of these, only 3 met this 
survey's primary selection criteria of $K\le$13.0 and $H-K \ge 1.67$. These three objects
(GY 12, IRS 26, and GY 312) are a subset of the 7 objects
listed in $\S$2 which met our selection criteria, but were unobserved by us.
Conversely, there are 4 sources meeting 
our selection criteria and not observed by us, which fell within
the areas of L1688 surveyed by {\it ISOCAM}, but which are not listed
among the 199 L1688 cloud members. These objects are GY30,
BKLT 162629-241748,  BKLT 162735-244628, and BKLT 162805-243354.
It is likely that the first 3 of these have mid-infrared colors 
of background objects, and were excluded from membership in
the L1688 cloud on that basis.  Note, however, that GY 30 has 
recently been discovered to drive a molecular outflow, and 
illuminates a fan-shaped reflection nebulosity \cite{kam03, den95}.
The source,  BKLT 162805-243354, just 4$^{\prime\prime}$ South of GY 472, may not have been
resolved with {\it ISOCAM}.  In summary, all of the objects meeting our selection
criteria, but unobserved by us, merit further investigation. 

Finally, 41 objects in Table 2, of which 37 fell within the ISOCAM survey's 
field of view, are not listed among the L1688 cloud members
observed by ISOCAM. The fact that only flux upper limits were measured
from the ground-based observations reported here for 34 of 37 of these objects, 
is consistent with their being background sources.
The three exceptions are GY263, GY232, and BKLT162618-242818, for which
there are ground-based detections (see Table 2), but no ISOCAM fluxes
are published.  Two of these, GY232 and BKLT162618$-$242818, have been
spectroscopically determined to be background giants \cite{luh99}.  The nature
of GY263 remains to be determined.
There are 4 objects in this ground-based survey which fell outside of
the {\it ISOCAM} survey's field-of-view:  WLY64=IRAS64a,  SR-2,
BKLT162904-244057,  and BKLT162522-243452. WLY64 exhibits an M8-M9 III optical spectral type, but a K4 III NIR CO-band absorption, very similar to FU Ori objects \cite{luh99}. SR-2 has a G8 spectral
type and had been assumed to be a foreground object based on its NIR colors \cite{eli78}.
More recent, high-dispersion, optical spectroscopy has shown SR-2 to
be a pre-main-sequence member of the $\rho$ Oph cloud core, however \cite{wal94}.
The nature of the other two objects remains to be determined.

\section{Discussion}

\subsection{Relative Timescales within SED Classes:  The Flat Spectrum Objects}

For nearly two decades now, young stellar objects have been
age-ordered according to an empirical classification scheme based
on the slope, $a$, of their near-infrared (2.2 $\mu$m) to mid-infrared (10 $\mu$m) spectral energy distributions (SEDs) :

$$a\ =\  {        d\, {\rm log}      \, (\lambda F_{\lambda})     \over d \, {\rm log} \, \lambda       }.$$
In this scheme, Class I SEDs have $a \ge 0.3$, Flat Spectrum SEDs
have $-0.3 \le a \le $+$ 0.3$, Class II SEDs have $-0.3 \ge a \ge -1.6$, and Class III SEDs
have $a \le -1.6$ \cite{gre94}. These latter authors 
were the first to suggest that the Flat Spectrum SEDs be identified as a distinct
class, since YSO's with $+0.3 \ge a \ge -0.3$ 
were found to have spectra strongly veiled
by continuum emission from hot, circumstellar dust, unlike the case for the 
near-infrared spectra of typical, classical T-Tauri stars (those generally associated
with Class II SEDs).  

In general, the above spectral slope classes have been found to correspond to distinct 
physical objects, defined by the amount and geometry of the circumstellar material surrounding
the central, forming YSO. Thus, a Class I object is one in which the central YSO has
essentially attained its initial main-sequence mass, but is still surrounded by a remnant infall
envelope and accretion disk; a Class II object  is
surrounded by an accretion disk; and a Class III object has only a remnant, or absent, accretion
disk. The gradual clearing of circumstellar matter has been interpreted as an evolutionary
sequence.  

In this context, in most cases the SED class reflects the evolutionary state.
As a cautionary note, however, there may be instances in which the SED 
class does not give the correct evolutionary state. For instance,
a T-Tauri star seen through a great amount of foreground obscuration may display a Class I or Flat
Spectrum SED (since the 2 $\mu$m flux suffers heavier extinction that the 10 $\mu$m flux).
Additionally, orientation effects can also be very important, since a more pole-on Class I object
may have an SED similar to an edge-on Class II object \cite{whi03}.  The definitive way
to ascertain the evolutionary state of a YSO is to obtain resolved images of the given object
at several different infrared wavelengths, and to compare these images quantitatively with 
model images produced using 3-D radiative transfer codes \cite{whi03}.  Since such data 
are unavailable for our current source sample, our analysis of the current dataset, in terms
of ascertaining the relative number of objects in each evolutionary state, is necessarily
limited to a plot of the distribution of spectral slope values, $a$.

Figure 4 shows histograms of the near- to mid-infrared spectral slope distributions
for the young stellar population of the  $\rho$ Ophiuchi cloud core. 
The solid line shows the spectral slope distribution determined
for the objects in our ground-based study. The spectral slopes, $a$, were
determined using the near-infrared (2.2 $\mu$m) photometry of BKLT 
and the mid-infrared photometry presented in Table 2 (at 10.8 $\mu$m for MIRLIN data
and  at 12.5 $\mu$m for the LWS data). The dashed line shows the 
spectral slope distribution of objects from the {\it ISOCAM} study.
Spectral slopes, $a$, in this case were determined from the 2.2 $\mu$m photometry
of BKLT and the 14.3 $\mu$m photometry from  {\it ISOCAM} \cite{bon01}.

It is striking that there are more objects exhibiting Flat Spectrum near- to mid-infrared
spectral slopes, $a$,  than Class I spectral slopes in Figure 4. 
The number of sources exhibiting Flat Spectrum slopes is
about 1/2 the number exhibiting Class II spectral slopes. 
This result holds for both the ground-based (solid line) the {\it ISOCAM} (dashed line) histograms.

Recent two-dimensional radiative transfer modelling has shown that the SEDs of 
Flat Spectrum objects are reproduced by a central star $+$ disk system embedded
in a relatively tenuous halo \cite{kik02}.  The dusty halo in Flat Spectrum objects is what
is left  of the remnant infall envelope with cavities carved out by outflows seen in the Class I objects.
The dusty halo of Flat Spectrum objects serves to heat the disk by scattering and reprocessing the central PMS star's  radiation.  It is the photosphere of the warmed disk that is responsible for the 
large mid- to far-infrared excesses that produce the observed flat spectral slope observed in these 
objects.

The reason previous authors have neglected to remark upon the large fraction of sources
exhibiting Flat Spectrum slopes
in the $\rho$ Oph cloud is because they had de-reddened the spectral slopes 
assuming these objects had intrinsic colors of T-Tauri stars (Bontemps et al. 2001, Wilking
et al. 2001).  Such a procedure relies on the assumption that the observed source is a Class II object,
seen through high foreground extinction.
However,  in the presence of heated and reprocessing halos,
such as are present in the Flat Spectrum phase, such a dereddening procedure is problematic.
Complications due to heating and re-processing of both stellar and disk radiation
in the remnant infall envelopes also preclude the ``de-reddening'' of Class I objects.

The important result here is that there is a non-negligible phase in YSO evolution
marking the transition between the Class I and Class II or Class I and young Class III phases.
During this reasonably lengthy transitional, or Flat Spectrum phase, a dusty halo remnant envelope remains and is slowly dispersed.  Of the sources in our survey that could be assigned
to SED classes based on our data, there are 19 Class I, 23 Flat Spectrum,
37 Class II, and 21 Class III objects. The {\it ISOCAM} data from Tables 2-5 of Bontemps et al. (2001)
contain 20 Class I, 30 Flat Spectrum, 78 Class II,  and 19 Class III objects (note, however, that the 
{\it ISOCAM} SED slopes are derived from 2--14 $\mu$m).  Flat Spectrum objects make up 
23\% and 20\% of the total number of classified objects in each survey, respectively.
The ratio of Flat Spectrum to Class I objects is 1.2  in our survey and 1.5 in the {\it ISOCAM}
survey.  The typical lifetime of an object in the Flat Spectrum phase is comparable (or
perhaps slightly greater than) that in the Class I phase.  For a solar-luminosity object,
this would correspond to 4 x 10$^5$ yr based on the study of Wilking et al. (1989).

\subsection{Active Accretion and Variability at Mid-Infrared  Wavelengths}
 
Table 3 lists the objects whose fluxes, as measured in this survey, differ significantly from
their published {\it ISOCAM} fluxes, or from other previously published ground-based 
mid-infrared photometry. Sources are included in Table 3
if the difference between two flux measurements
exceeds the quoted 3-$\sigma$ errors. To compile this table,
we have included previously published mid-infrared
photometry, in addition to the fluxes
plotted in Figures 2 \& 3 \cite{wil01,gre94,wly89,lad84,ryd76}.
In Table 3, the first column lists the BKLT source designation, and the second
column lists a more common alias.
The 10.8 $\mu$m flux and its associated error measured
with MIRLIN are listed in the third colum.
For direct comparison, the interpolated 10.8 $\mu$m flux
and its associated error from the {\it ISOCAM} data are listed in
the fourth column.  The 12.5 $\mu$m fluxes measured with LWS are
next listed in the fifth column, and the interpolated 12.5 $\mu$m fluxes
and their associated errors from {\it ISOCAM} are listed in the sixth column.
Finally, the last column lists previous ground-based mid-infrared photometry.
These latter fluxes are all at
10.2 $\mu$m, except for WL17, whose 12.5 $\mu$m flux is listed.

The source of the flux discrepancies in Table 3 can either be intrinsic source
variability or the presence of extended emission.  Intrinsically variable sources
in Table 3 are those for which fluxes from this
survey exceed measured fluxes from previous surveys (all of which used larger apertures).
For sources whose previous flux measurements, taken in larger apertures, exceed the fluxes
from this survey, further high-resolution photometry is required to distinguish between
source variability and the presence of significant extended emission. These sources 
are, nevertheless, included in Table 3 as possibly variable sources.

Figure 5 shows the SED slope distribution of the mid-infrared
variables and candidate variables of Table 3.  Clearly, mid-infrared variability occurs for all SED
classes with optically thick disks. We have also
plotted in Figure 5 the SED slope distribution of known near-infrared variables
in $\rho$ Oph from Table 5 of Barsony et al. (1997).
Near-infrared variability, also found for all SED classes with optically
thick disks, seems to occur with somewhat higher frequency at the earlier evolutionary stages
(Flat Spectrum and Class I).  Whereas 89\% of NIR variables are in the Class I or Flat Spectrum
stage, only 56\% of the mid-IR variables are.  This tendency must be verified from future
sytematic NIR and mid-IR variability studies to improve the statistics.

Sources exhibiting variability at both near- and mid-infrared wavelengths are
WL12, WL17, WL19, GY244, GY245, GY262,  and IRS44.  
WL15, an almost face-on Class I object,  appears variable at near-, but not at mid-infrared
wavelengths.  The rest of the mid-infrared variables listed in Table 3 are apparently
not highly variable at near-infrared wavelengths. However, further systematic variability
studies in the infrared are needed to verify these preliminary results.

In order to understand what distinguishes the population of 
mid-infrared variables from the rest of the embedded population
in $\rho$ Oph, we use the data compiled in Table 4 to look
for a correlation between mid-infrared variability and 
presence of an active accretion disk, as signalled by
near-infrared veiling.  The $r_K$ values tabulated in Table 4
are all {\it spectroscopically} determined, and are, therefore,
reddening-independent.  By using moderate ($R\sim 2000$)
to high ($R\sim 20000$) resolution near-infrared spectroscopy, 
one can deduce the $K-$band ``veiling''
of photospheric absorption lines in a YSO by comparison with the shapes of
the same absorption lines found in a spectroscopic standard star of the same spectral
type. One then varies the ``veiling'' applied to the standard star's spectrum
until the ``veiled'' standard star's spectrum best matches the YSO's spectrum.
Such a ``veiling'' measurement is independent of reddening,
given the narrow wavelength range in which the relevant spectral lines lie.
 
The objects in Table 4 are listed
in reverse near- to mid-infrared spectral slope ($a$) order,  from the largest values,
corresponding to Class I objects, to the smallest values, corresponding
to Class III objects. The spectral slopes in Table 4 were deterrnined using the 
near-infrared ($K$-band, or 2.2 $\mu$m) photometry of BKLT and either the
10.8 $\mu$m flux from MIRLIN, or the 12.5 $\mu$m flux from LWS, as available,
and as listed in Table 2.

We have excluded the known background giants from the entries in Table 4.
Also excluded are the two known PAH/VSG emitters, WL16 and WL22.
Table 4 lists the BKLT source designation in the first column and
an alternate alias for each source in the second column.
The K magnitude and H$-$K color for each object are listed in the third and 
fourth columns, respectively, and are taken from
Barsony et al. (1997), except for the following:  WL20, GSS30 IRS3 \cite{sks95}, 
GY197 \cite{crb93}, WL1 \cite{hai02}, WL18, Elias 26 \cite{hai03},  and DoAr24E \cite{che88}.
Each object's spectral slope, $a$, between 2.2 $\mu$m and either 10.8 $\mu$m or 12.5 $\mu$m
(as available from Table 2),
and corresponding SED class, are listed in the fifth and sixth columns respectively.
Published, spectroscopically-determined,  K-band veiling values, and visual extinctions are listed in the seventh and last columns, respectively. 

Continuum veiling at K (2.2 $\mu$m) is quantified by the value, $r_K$, 
defined as $F_{excess}/F_{K}$, where $F_K$ is the intrinsic photospheric flux of the 
central object at $K$, and $F_{excess}$ is the amount of observed $K$-band flux
in excess of the expected photospheric value.   Veilings of $r_K \ge 0.5$ have been attributed to the presence of actively accreting circumstellar disks \cite{luh99,wil01}. Figure 6 presents
a graph of the  near- to mid-infrared spectral slopes, $a$, from Table 4, plotted
against the {\it spectroscopically determined} $r_K$ values published by various authors 
(Greene \& Meyer 1995; Wilking et al. 1999; Luhman \& Rieke 1999; Barsony et al. 2002; Greene \& Lada 2002; Prato et al. 2003; Doppmann et al. 2003; Doppmann 2004).
The horizontal dashed line in Figure 6 denotes
the boundary between optically thick (above the line) and optically thin (below the line)
disks, at  $r_K = 0.58$ \cite{wil01}.   Class I  objects generally
exhibit $r_K \ge$ 1 (the one exception being GY91/CRBR42 with $r_K=0.3$).
Flat Spectrum objects  generally have $r_K \ge 0.58$ (with the exception of 
GY244, with $r_K=0$). Class II objects have the
widest range of $r_K$ values, ranging from $0 \le r_K \le 4.5$ (the record of 4.5 being
held by Elias 26). Class III objects, clearly have optically thin disks, most with $r_K=0$. 
The most striking feature of this graph  is the highly variable value of veiling
that a single source can exhibit in any of the SED classes in which active disk
accretion can take place ({\it i.e.}, in the Class II/Flat Spectrum/Class I phases).
This is direct observational evidence for highly time-variable accretion activity in disks.

\subsection{Disk Structures}

A current problem of great interest is how disks are dissipated in YSOs.
Is it possible that disk winds would preferentially clear the innermost
portions of a disk before the outer portions, or vice versa? One
possible way to address this question is to look for the 
frequency of near-infrared vs. mid-infrared excesses, perhaps as a
function of the age of the system,  since near-IR excesses
originate from disk regions much closer to the central object than mid-IR excesses.

For this purpose, we will consider only those Class II and Class III objects
whose effective temperatures have been spectroscopically determined, and that have
good estimates of A$_V$.  We have omitted the known binaries, S1, SR24N, WL13, Elias 34, IRS2, and GSS29 \cite{bar03,dop04} from this plot. These systems would either be unresolved by
our mid-infrared data (spectroscopic binaries) or their projected angular separations
are so small that both of their spectra would fall into a single slit, thereby invalidating
the derived effective temperature and visual extinction values. Individual components
of well-resolved multiple systems  have been included in Figure 7, however.

We calculate $K$-band excesses, $\Delta K$, and $N$-band excesses,
$\Delta N$, defined as the logarithm of the ratio of the 
observed flux to the flux expected from the PMS star's photosphere at 2.2 $\mu$m
and 10.8 $\mu$m, respectively, seen through the appropriate amount of 
foreground extinction. More specifically, we use the definitions, 
$\Delta K = log{ F(observed)_{2.2\mu m}/F(photospheric)_{2.2 \mu m} }$
\cite{str89}, and 
$\Delta N = log{ F(observed)_{10.8\mu m}/F(photospheric)_{10.8 \mu m} }$
(Skrutskie et al. 1990). Note that this method of calculating $r_K$ is reddening
dependent, since it results in a different $\Delta K$ value for a different input $A_V$,
in contrast to the spectroscopically determined values of $r_K$ used in Figure 6.

Figure 7 shows the resulting $\Delta N$ vs. $\Delta K$ plot.  In the left panel,
the expected photospheric contributions at K and N were determined assuming
a blackbody radiator at the spectroscopically determined effective temperature for
each source. In the right panel, Kurucz model atmospheres of the appropriate
temperature, with solar abundances, and $log\ g = 3.5$,
were used to infer the photospheric emissions at K and N.
Class III objects are plotted as triangles, and Class II objects as squares.
The demarcations between opticallly thin/thick disks 
at $\Delta K=$0.2 and at $\Delta N=$1.2 are indicated by the dashed and dotted lines, 
respectively \cite{skr90}.

Class III objects, as expected, have optically thin disks at both near- and mid-infrared wavelengths,
with $\Delta K$ values scattered about 0.0 (for the blackbody models).  There is an inherent bias in the sample  of  Class III objects plotted in Figure 7, since  in order to appear on this plot, they are all detected in the  mid-infrared.  Therefore, instead of being scattered about a mean $\Delta N = 0.0$, there
appears to be a positive bias in the $\Delta N$ values of these Class III objects.

The sample of Class III
objects plotted here are unique in that they have any detectable mid-infrared emission
above photospheric at all, signalling the presence of 
surrounding disks, even though these are opticallly thin. 
These seem to have been cleared possibly from inside out 
($\Delta$ K consistent with photospheric
emission only, but $\Delta$ N above photospheric).  

The dramatic finding in Figure 7 is the lack of any objects occupying
the lower right quadrant of each plot, corresponding to optically thick inner disks
with optically thin outer disks, for both the blackbody and Kurucz model
central objects.  This finding is consistent with two possible disk dissipation
scenarios:  either disk clearing proceeding at the same rate at all radii or disk
clearing from the inside-out.  For the case of disk clearing proceeding at the
same rate at all radii, we would expect to see a clear correlation between $\Delta$ K
and $\Delta$ N, whereas for disk clearing from the inside-out, we would expect
all of the quadrants of Figure 7, except the lower right, to be significantly populated.

At first sight, it may seem as though there is a correlation between $\Delta$ K
and $\Delta$ N, in the left-panel of Figure 7, with the exception of GY11.
However,  the sources, GY314, GSS29, GSS39 have been previously pointed out as examples of disks with inner holes \cite{wil01}, which should occupy the upper left quadrant in each panel
of Figure 7.  In fact,  GY314 does lie in the upper left quadrant of each panel of Figure 7, whereas
GSS 39 lies no further than $\Delta$ N$=$ 0.1 from this quadrant even in the case of a blackbody model for the central object (GSS29 is a newly
discovered spectroscopic binary, and has been excluded from Figure 7 on that basis).
If we allow an error of just $\Delta$ N $=$0.1 in the mid-IR excess determinations
for the blackbody models, then the upper two quadrants for both central source models would be equally occupied, and one could argue for evidence for disk clearing from the inside-out.

Of course, were we to consider only the right panel, where Kurucz model atmospheres have been used to model the central objects, we would conclude that disk clearing does proceed from the inside out.
However, we must point out a systematic error that seems to be present in the near-infrared excesses
as determined when using Kurucz model atmospheres for the central objects:  Instead of being
equally distributed about a $\Delta$ K $=$ 0 value, as would be expected for random errors for
lack of NIR excess emission, the objects with optically thin disks are scattered about a mean
negative value of $\Delta$ K $= -$0.1. This systematic error could be caused by the presence of
just 10\%--20\% veiling at J band, or by the presence of some scattered light at J.

To definitively decide on  the disk clearing mechanism, two avenues of further investigations
are called for. First,
spectral type determinations for more Class II and Class III sources are needed,
in particular, ones for which the K and N band fluxes are known, in order to improve the statistics
of Figure 7.  Secondly, detailed investigations of the contributions to each individual object's
J band flux from scattering and/or veiling are needed, in order to more accurately determine the
K band excesses when using Kurucz model atmospheres.

%

\subsection{Effect of Multiplicity on Disk Evolution}

Table 5 lists the targets observed in our mid-infrared survey
that are known to be in multiple systems (Barsony, Koresko, \& Matthews 2003). 
Of a total of 22 multiple systems observed,
two were new discoveries:  IRS 34, at 0.31$^{\prime\prime}$ separation, and WL1
at 0.82$^{\prime\prime}$ separation. Fifteen of the observed, known multiple
systems were resolved in this study. Individual component fluxes and/or flux
upper limits are listed in Table 5.

Data from this survey have been used to study the multiplicity
frequency among Class I protostars, with the result that
the restricted companion fraction for the observed magnitude difference
and separation range is consistent with that found for Class II objects.  Thus,
the restricted companion fraction for both Class I and Class II objects
exceeds that found for main-sequence stars by a factor of two \cite{hai02,hai04}.
These authors also inferred that many sub-arcsecond Class I binaries remain to be 
discovered. 

The question of how binarity affects disk evolution may be divided into 
several categories, based on binary separation \cite{loo00}.
Gravitationally bound objects evolving from separate envelopes ($\ge$ 6500 AU separation),
objects with separate disks, but common envelopes (100 AU $\le$ r $\le$ 6500 AU),
and objects sharing both a single disk and a single envelope ($\le$ 100 AU separation)
must evolve differently. The angular scales accessible in this survey
are most amenable for studies of multiple systems that have separate disks, but common-envelopes.

A detailed study of an unusual triple system, WL20, that was part of this survey
has shown that disk-disk interaction has resulted in enhanced accretion 
onto one component of this system, WL20S.  This tidally-induced disk disturbance
explains the Class I SED of this object, although it is coeval, at an age of 
several million years, with its Class II SED companions \cite{res01,bar02}.

A comprehensive study of disk evolution in these multiple systems requires both
spatially resolved, multi-wavelength, near-infrared photometry--a challenge at 
sub-arcsecond separations and at the faint J magnitudes of many of these systems,
and spatially resolved near-infrared spectroscopy.  Such data have been obtained
for a dozen binary systems and their analysis is the subject of a future work \cite{bar05}.

\section{Summary}

We have carried out a high-spatial resolution, ground-based, mid-infrared imaging survey
of 172 objects toward the $\rho$ Ophiuchi star-forming cloud core. The target list
included 102 objects chosen for their combined properties of infrared brightness
(K$\le$13.0) and red color (H$-$K$\ge$1.67) from the NIR survey of Barsony et al. (1997),
augmented by known cloud members inferred from observations at other wavelengths.

Eighty-five of the target objects were detected. The general agreement between the
mid-infrared fluxes determined in this study with the {\it ISOCAM}-determined fluxes
for objects common to both surveys is quite good, as is the agreement for objects
in common to both studies for which only ground-based flux upper limits could be determined.
A significant fraction of sources (18 objects) were found to be mid-infrared variables,
and a further 19 objects are either extended and/or variable.

A plot of the frequency near- to mid-infrared spectral slopes, $a$, for the
objects with newly determined mid-infrared fluxes, shows 19 Class I, 23 Flat Spectrum,
37 Class II, and 21 Class III objects.  It is argued that Flat Spectrum objects
represent a distinct evolutionary phase in which the remnant infall envelopes
from the Class I phase are dispersed, and, that YSOs spend a significant 
fraction of time, of order 4 $\times$ 10$^5$ yr, in this state.

A plot of the spectroscopically determined near-infrared veiling,
$r_K$  vs. near- to mid-infrared spectral slope, $a$, for the detected objects in our
survey for which published $r_K$ values are available, is presented. 
We find a general trend of  an increasing $r_K$ {\it threshold} with increasing $a$,
such that  Class I  objects {\it generally}
exhibit $r_K \ge$ 1, Flat Spectrum objects {\it generally} have $r_K \ge 0.58$,
and Class III objects {\it generally} have $r_K=0$. 
Class II objects, however, have the
widest range of $r_K$ values, from $0 \le r_K \le 4.5$.
The most striking result, however, is the highly variable value of veiling
that a single source can exhibit in any of the Class II/Flat Spectrum/Class I phases,
signalling the highly time-variable accretion activity in disks.

Finally, to study disk dispersal mechanisms, we present plots of $\Delta$N (mid-IR excess
above photospheric emission) vs. $\Delta$ K (NIR excess above photospheric emission),
for two photospheric models:   blackbody and Kurucz model atmospheres.
Determinations of the mid-IR and NIR excesses were made  
for the subset of our mid-IR survey sample for which spectroscopically determined effective temperatures
and reliable, previously published, $A_V$ values are available.   
In all cases, we find no sources occupying the region of the plot that corresponds to optically
thick inner disks with optically thin outer disks.  By contrast, the entire region of optically thin inner disks,
spanning the range from optically thin to optically thick outer disks, is populated, as is the region
with both optically thick inner and outer disks.  The results are consistent with disk dispersal proceeding from the inside-out, but further observational investigations to confirm this hypothesis
are suggested.

\acknowledgements
This paper is dedicated to the memory of a dear friend, colleague, and mid-infrared
astronomer extraordinaire, Dr. Lynne K. Deutsch, who passed away after a long illness
on the night of 2 April 2004.

Financial support through NSF grants AST 00-96087 (CAREER), AST 97-53229 (POWRE), 
AST 02-06146, the NASA/ASEE Summer Faculty Fellowship program at the Jet Propulsion Laboratory, and NASA's  Faculty Fellowship Program at Ames Research Center, have made 
the completion of this work possible.  We wish to thank the staff at the Keck and Palomar observatories
for their enthusiasm, patience, and assistance in making it possible to use MIRLIN at these telescopes,
and Bob Goodrich for his support with the LWS instrument at Keck I.  Development of MIRLIN was supported by the JPL's Director's Discretionary Fund, and its continued operation was funded by an
SR$+$T Award from NASA's Office of Space Science.

Dr. Suzanne Casement, former post-doctoral researcher at U.C. Riverside, Ana Matkovic,
former summer undergraduate researcher at Harvey Mudd College, and former HMC undergraduate, Christian Baude, also contributed during the early data reduction phases of this work. Patricia Monger's contributions in systems administration and in providing the plotting program, SM, have been invaluable, and are gratefully acknowledged. We thank the referee, Prof. Bruce Wilking, for his meticulous
readings of the manuscript, and for his numerous helpful suggestions for its improvement.
\clearpage
\clearpage
.
\enddata
\tablenotetext{1}{Flux from Table 2 of  Wilking, Lada, \& Young (1989)}
\tablenotetext{2}{Flux from Table 4 of Rydgren, Strom, \& Strom (1976)}
\tablenotetext{3}{Flux from Table 1 of Lada \& Wilking (1984)}
\tablenotetext{4}{Flux from Table 2 of Elias (1978)}
\tablenotetext{5}{Flux from Table 1 of Greene et al. (1994)}
\tablenotetext{6}{The SR24 system is a hierarchical triple, with a 6.0$^{\prime\prime}$ separation
between the primary, SR24S ($=$BKLT J162658$-$244534), and the secondary, SR24N ($=$ BKLT J162658$-$244529),
which itself consists of a 0.20$^{\prime\prime}$ separation binary. Although SR24S \& SR24N were unresolved by {\it ISOCAM}, 
previous ground-based observations did resolve these components.  N-band fluxes for SR24S are  0.78 $\pm$ 0.029 Jy (WLY89) and
2.28$\pm$0.21 Jy (GWAYL94), to be compared with MIRLIN fluxes of 2.01 Jy and 1.66 Jy for SR24S on 26 June 1997 and 7 June 1998,
respectively. N-band fluxes for SR24N are 1.17 $\pm$ 0.029 Jy (WLY89) and 1.67$\pm$0.16 Jy (Greene et al. 1994), to be compared with the MIRLIN flux
of 1.40 Jy for both dates of observation.  Therefore, we can conclude
that it is the single, primary star in this system, SR24S, which is the mid-infrared variable.}
\tablenotetext{7}{WL20 ($=$ BKLT J162715$-$243843) is a triple system, which was noted as highly
variable in unresolved {\it ISOCAM} 6.7 $\mu$m measurements. The previously published N-band
flux for the WL20 system from ground-based measurements was 0.180 $\pm$ 0.03 Jy (LW84).
Our observations resolve the triple system in the mid-infrared,
and show that it is the ``infrared companion'', WL20S, 
which is the mid-infrared variable (see Table 2).}
\tablenotetext{8}{{\it ISOCAM} data from Wilking et al. (2001).}
%
%
\end{deluxetable}
\begin{deluxetable}{llcccccc}
\rotate
\tablecolumns{8}
\tablewidth{0pc}
\tabletypesize{\small}
\tablecaption{$\rho$ Oph Source Properties}
\tablenum{4}
\tablehead{ 
\colhead{BKLT} & \colhead{Aliases} & \colhead{K} & \colhead{H$-$K}& \colhead{$\alpha_{IR}^{1}$} & \colhead{SED} 
& \colhead{$r_K^2$} 
& \colhead{$A_V^2$} \\
\colhead{}           & \colhead{}              & \colhead{(mag)} & \colhead{(mag)} & \colhead{}                                      &  \colhead{}         &  \colhead{}          & \colhead{(mag)} 
}
\startdata
162722$-$242939     &GY256                              & 12.75 & 2.79                        & $+$2.35  &    I   & \nodata     &  \nodata      \\
162751$-$243145      &IRS54                               & 10.87  & 2.63                       & $+$1.77   &   I   &  $>$2         & \nodata   \\
162724$-$244103     &CRBR85                           &13.21   & $\ge$ 2.29            & $+$1.59   &   I   &  \nodata    & \nodata    \\
162715$-$243843      &WL20                                &             &                                &                   &        &                   &     \\
                                        &WL20S                               & 12.57  & 1.88                    &$+$1.38    &  I    &  $<0.9$ (BGB02)    & 41.3 (BGB02)   \\
162709$-$243718      &WL15                                  & 7.54    & 4.47                    & $+$1.27   &  I    &  $>2$       & 11.1 $\le A_V \le$ 17.0 (B02)  \\
162729$-$243917     &IRS46/YLW16B                  & 11.46  & 3.19                    &$+$1.15   &  I    &  $>1$    &  \nodata     \\
162617$-$242347    &CRBR12                              & 12.04  & 3.54                    & $+$1.15   &  I    & 2.0 (D04) & \nodata     \\
\end{deluxetable}.
%
\enddata
\tablenotetext{1}{$\alpha_{IR}$ is calculated between 2.2 $\mu$m and either 10.8 $\mu$m or 12.5 $\mu$m, as available from Table 2.}
\tablenotetext{2}{$A_V$ and $r_K$ values are from LR99, unless otherwise indicated.}
\tablenotetext{3}{The nature of WLY64 remains to be clarified, although it  is listed as an FU Ori candidate source by LR99.}
\tablenotetext{4}{Spectroscopically determined background giant (Luhman \& Rieke 1999).}
\tablerefs{BKLT $=$ Barsony et al. 1997; B02 $=$ Boogert et al. (2002); BA92 $=$ Bouvier \& Appenzeller 1992; 
 BGB02 $=$ Barsony, Greene, \& Blake 2002;  C88 $=$ Chelli et al. 1988; DJW03 $=$ Doppmann, Jaffe, \& White 2003;
 D04 $=$ Doppmann 2004; GL97 $=$ Greene \& Lada 1997;  GL02 $=$ Greene \& Lada 2002; GM95 $=$ Greene \& Meyer 1995;  
 HBGR02 $=$ Haisch et al. 2002;  LR99$=$ Luhman \& Rieke 1999;  PGS03 $=$ Prato, Greene, \& Simon 2003; 
 SKS95 $=$ Strom, Kepner, \& Strom 1995;  W94 $=$ Walter et al. 1994; W01 $=$ Wilking et al. 2001; WGM99 $=$ Wilking, Greene, \& Meyer 1999}
\end{deluxetable}
\clearpage
\begin{deluxetable}{llccrccr}
\rotate
\tabletypesize{\scriptsize}
\tablewidth{0pt}
\tablenum{5}
\tablecolumns{8}
\tablecaption{Mid-Infrared Observations of Known Binary/Multiple Systems in Ophiuchus}
\tablehead{
\colhead{BKLT} &
\colhead{Alias} &
 \colhead{$\alpha$(2000)} &
 \colhead{$\delta$(2000)} &
 \colhead{System} &
 \colhead{Sep'n.} & 
 \colhead{P.A.}   & 
 \colhead{Resolved in} \\
 \colhead{}     & 
 \colhead{}     & 
 \colhead{}     & 
 \colhead{}     &
 \colhead{Type\tablenotemark{1}} & 
 \colhead{Arcsec}  & 
 \colhead{Deg\tablenotemark{2}} & 
 \colhead{Mid-IR?}}
\startdata
                                   & ROX 1                    &16 25 19.28&-24 26 52.1&B   &0.236           &156         &N     \\
162536$-$241544 & IRS 2                      &16 25 36.75&-24 15 42.1&B   &0.44+/-0.03 &79+/-4  &N   \\
162623$-$242101 & DoAr 24E               &16 26 23.38&-24 20 59.7&B   &2.05             &148.6    &Y   \\
                                   & Elias23$+$GY21  &                      &                    &B &10.47             &322.6    & Y   \\
162624$-$242449 & Elias 23                  &16 26 24.06 &-24 24 48.1&   &                       &               &   \\
162623$-$242441 & GY 21                     &16 26 23.60 &-24 24 39.4&   &                       &               &   \\
162630$-$242258 &VSSG 27                 &16 26 30.50&-24 22 57.1 &B &1.22+/-0.03  &68 +/-1 &Y   \\
162634$-$242330 &S1                             &16 26 34.18&-24 23 28.2 &B &0.020            &110     &N  \\
162642$-$242031 &GSS37                     &16 26 42.87&-24 20 29.8  &B(T?) &1.44      &67.0    &Y   \\
162646$-$241203 &VSS27                     &16 26 46.44&-24 12 00.0  &B        &0.59       &104.6   &N  \\
                                   &WL 2                        &                      &                       &B       &4.17       &343       &Y \\
162648$-$242840 &WL 2(A)                   &16 26 48.50& -24 28 38.7  &         &               &              &   \\
162648$-$242836 &WL 2(B)                   &16 26 48.42& -24 28 34.7  &         &               &              &   \\
162649$-$243823 &WL 18                      &16 26 48.99& -24 38 25.1   &B      &3.55      & 293    &Y  \\
                                   &SR 24                      &                      &                         &T       &5.093   &60        &Y \\
162658$-$244534 &SR 24A                   &16 26 58.52 &-24 45 36.7    &         &              &            &    \\
162658$-$244529 &SR 24B                   &16 26 58.45 &-24 45 31.7    &         &0.197    & 84      &N \\ 
162704$-$242830 &WL 1                        &16 27 04.12 & -24 28 29.9  &B       &0.82      & 321.2  &Y \\
                                   &SR21                       &                      &                        &B       &6.33       &175     &Y \\
162710$-$241914 &SR 21A                   &16 27 10.28&-24 19 12.6    &          &               &            &   \\
162710$-$241921 &SR 21B                   &16 27 10.33&-24 19 18.9    &          &               &            &   \\
162715$-$242640 &IRS34                      &16 27 15.48&-24 26 40.6    &B       &0.31       &236     &  \\
162715$-$243843 &WL20                       &                     &                         &T        &              &            & \\
                                   &WL20W                   &16 27 15.69&-24 38 43.4   &           & 3.17 (E-W sep'n.)  &270     &Y \\ 
                                   &WL20S                    &16 27 15.72&-24 38 45.6   &           & 2.26 (S-W sep'n. )  &173     &Y \\
                                   &WL20E                    &16 27 15.89&-24 38 43.4   &           & 3.66  (S-E sep'n. )  &232     &Y \\ 
                                   &SR12$+$IRS42    &                     &                        &T          &               &            &   \\
162719$-$244139 &SR12                      &16 27 19.55&-24 41 40.0   &B         &0.30       &85        &N \\
162721$-$244142 &IRS42                     &16 27 21.45&-24 41 42.8   &            &26.8      &85.8     &Y \\
                                   &GY263$+$IRS43  &                     &                        &B         &6.99       &322    &Y   \\
162726$-$244045 &GY263                    &16 27 26.63&-24 40 44.9    &          &               &           &    \\
162726$-$244051 &IRS 43                    &16 27 26.94&-24 40 50.      &           &               &          &     \\
                                   &IRS 44$+$GY262&                      &                         &T(?)   &                &         &Y    \\
162726$-$243923 &GY262                    &16 27 26.49&-24 39 23.0    &           &23.21     &         &      \\
162728$-$243934 &IRS 44                    &16 27 28.01&-24 39 33.6    &B         &0.27       &81    &N    \\
162727$-$243116 &WL 13                     &16 27 27.40& -24 31 16.6   &B        &0.46       &356   &N    \\
162730$-$242744 &VSSG 17                &16 27 30.17&-24 27 43.5    &B        &0.25       &26     & N   \\
162740$-$242205 &SR9                         &16 27 40.28&-24 22 04.3    &B        &0.59       &350    &N    \\
162752$-$244049 &ROXs 31                 &16 27 52.07&-24 40 50.4   &B         &0.39       &71.6   &N   \\
%
%
\enddata
\tablenotetext{1}{B$=$Binary; SpB$=$Spectroscopic Binary; T$=$Triple; Q$=$Quadruple}
\tablenotetext{2}{PAs are E of N, measured from the primary at K, except for IRS34, where the primary is at N, since no resolved K band data exist}
%
%
\end{deluxetable}

\clearpage
%
%
\begin{figure}
\plotone{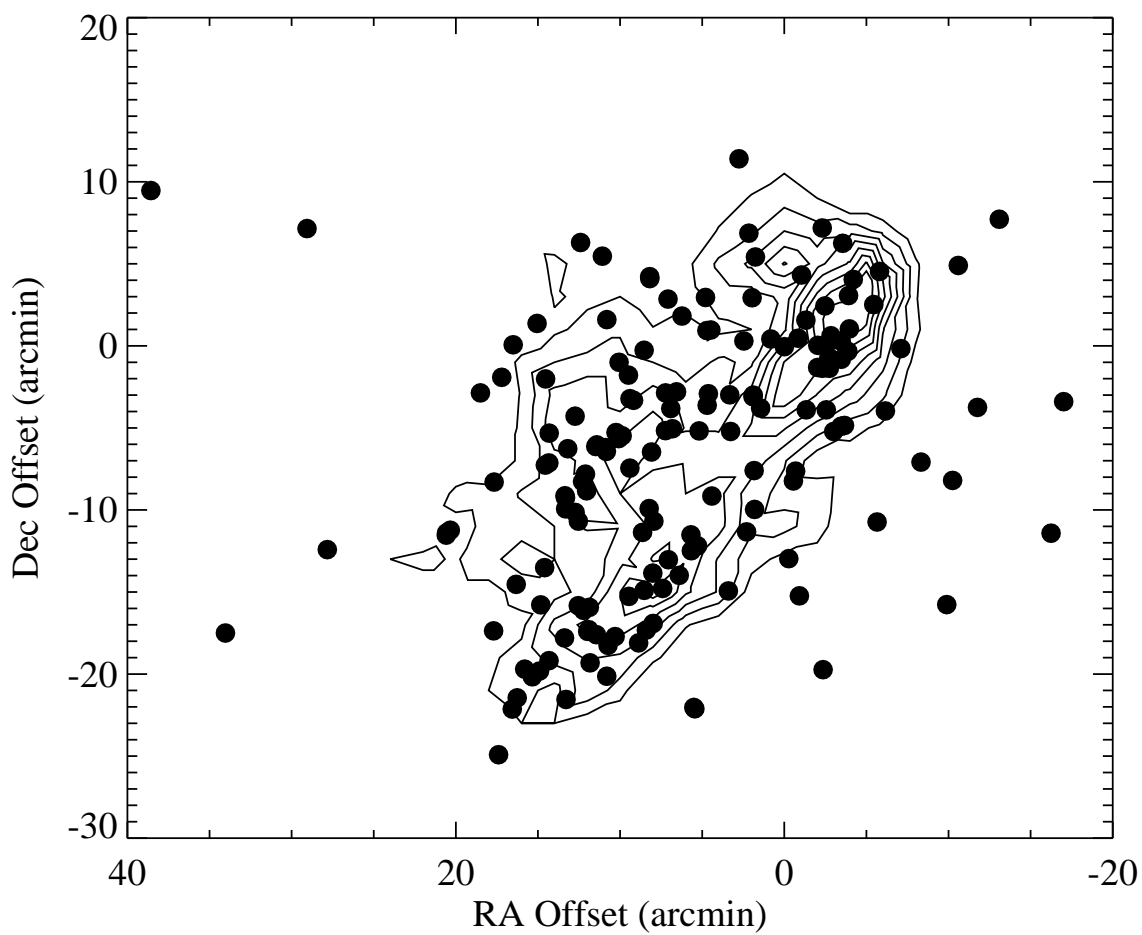}
\figcaption[f1.eps]{The spatial distribution of the target objects for which mid-infrared fluxes and/or flux 
upper limits are listed in Table 2.  The contour plot (courtesy of B. A. Wilking) shows the C$^{18}$O J$=$(2$\rightarrow$1) integrated 
intensity in K km/sec units, at 1 K km/sec intervals, starting at 3 K km/sec and ending at 11 K km/sec.  The offsets (in arcmin units) 
are from the (0,0) position of S1:
$\alpha_{2000}=$16h 26m 34.2s, $\delta_{2000}=-$24$^{\circ}$ 23$^{\prime}$ 27$^{\prime\prime}$.}
\label{Figure 1}
\end{figure}
\clearpage
\begin{figure}
\plotone{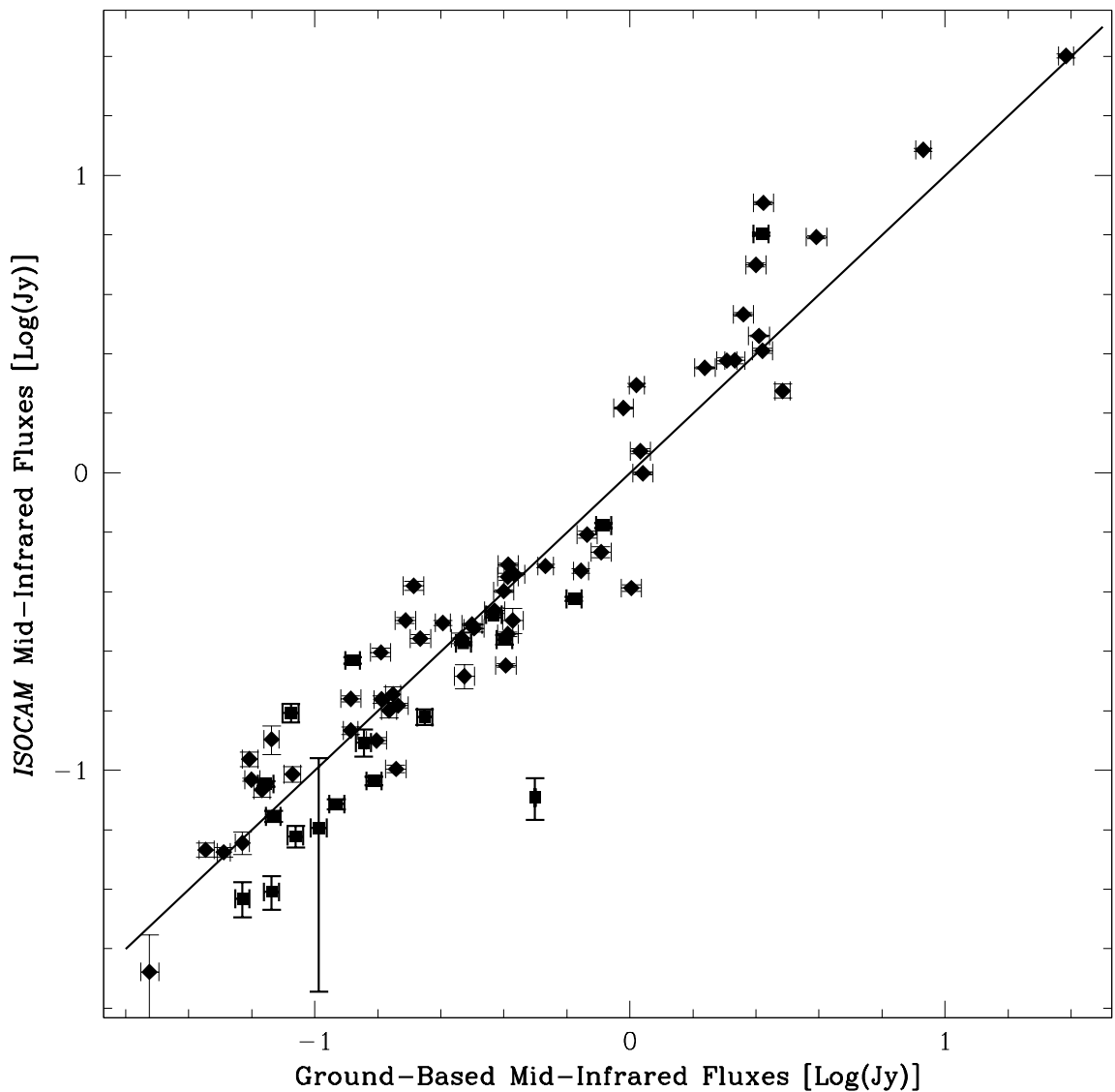}
\figcaption[f2.ps]{Plot of ground-based mid-infrared fluxes for objects in Table 2 vs. mid-infrared fluxes from {\it ISOCAM}. The 69 objects plotted here were common to, and detected by, both surveys.
Furthermore, each object plotted here was detected through both {\it ISOCAM} filters, centered at  6.7 $\mu$m and 14.3 $\mu$m, respectively.
Filled diamonds represent objects detected by MIRLIN at 10.8 $\mu$m, with 
linearly interpolated {\it ISOCAM} 10.8 $\mu$m fluxes.
Filled squares represent objects detected by LWS at 12.5 $\mu$m, 
with linearly interpolated {\it ISOCAM} 12.5 $\mu$m  fluxes.}
\label{Figure 2}
\end{figure}
\clearpage
\begin{figure}
\plotone{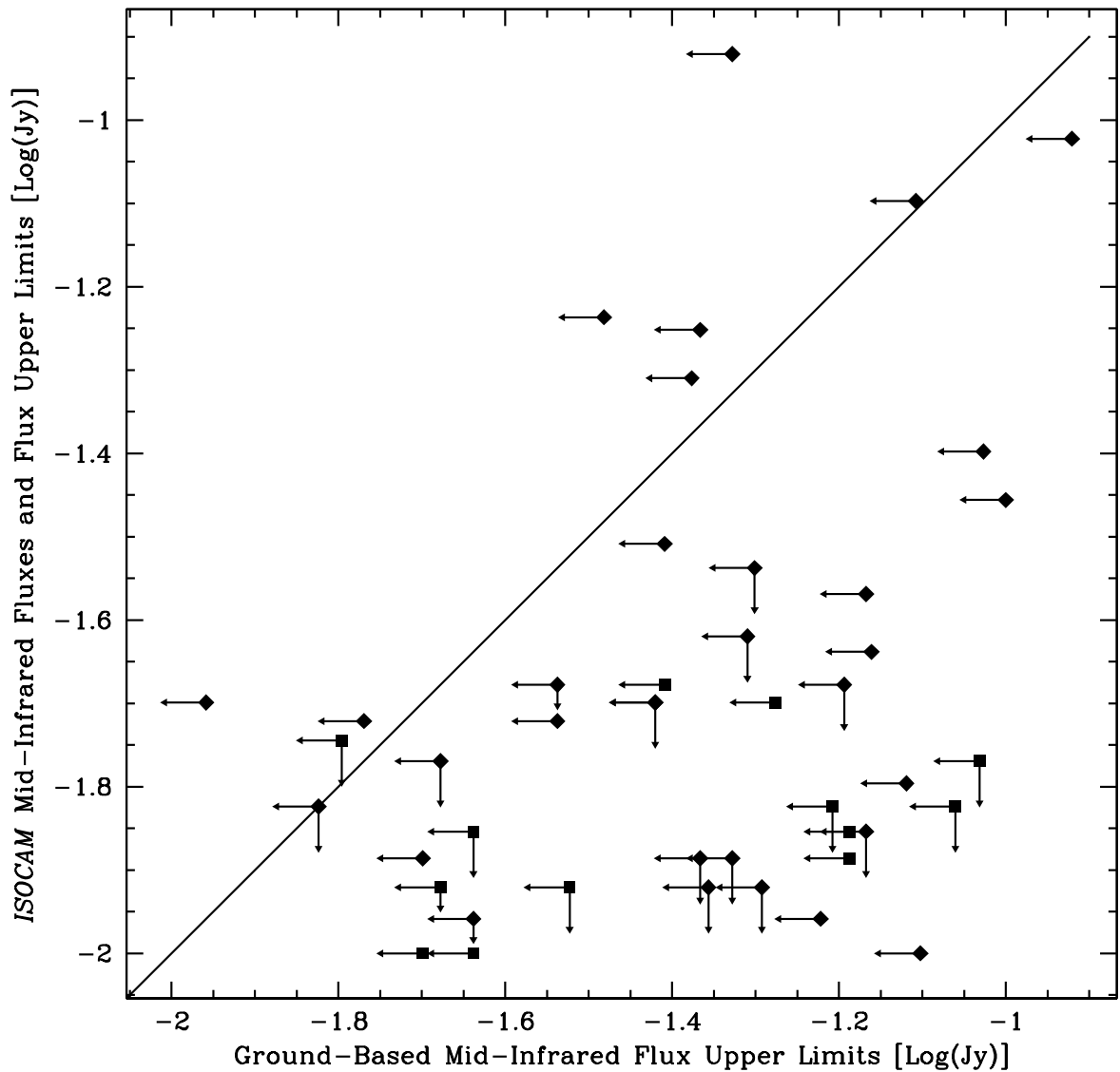}
\figcaption[f3.ps]{Plot of ground-based mid-infrared flux upper limits for objects in Table 2 in common with {\it ISOCAM}-detected objects. Filled diamonds represent MIRLIN 10.8 $\mu$m upper limits;
filled squares represent LWS 12.5 $\mu$m upper limits. {\it ISOCAM} fluxes were interpolated
to 10.8 $\mu$m (filled diamonds) or to 12.5 $\mu$m (filled squares) for objects detected in
both {\it ISOCAM} filters. For objects detected only at 6.7 $\mu$m by {\it ISOCAM}, 
the 14.3 $\mu$m completeness limit of 15 mJy is used to derive the interpolated {\it ISOCAM} flux upper limits.}
\label{Figure 3}
\end{figure}
\clearpage
\begin{figure}
\plotone{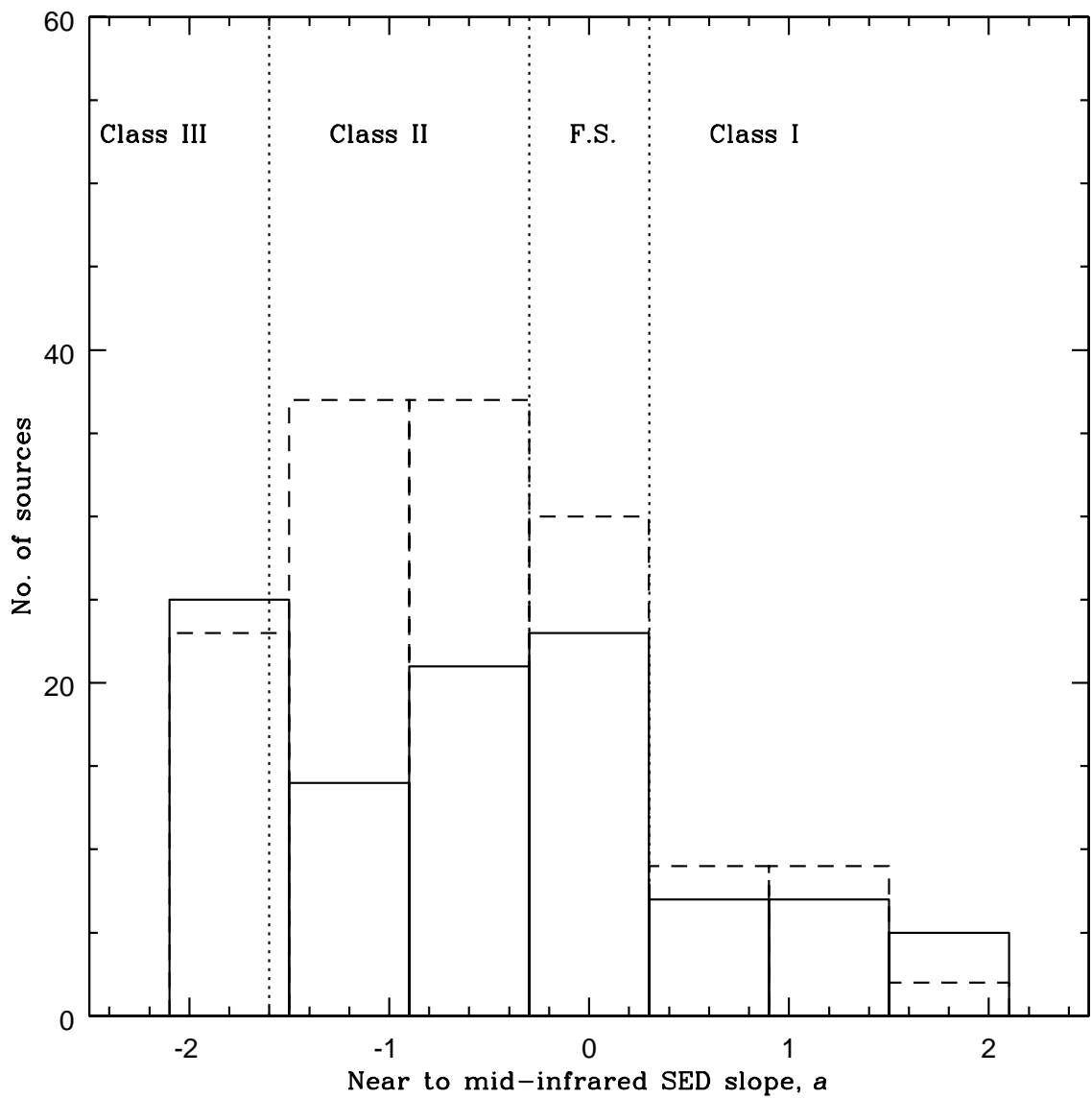}
\figcaption[f4.ps]{Histogram of the {\it observed} distribution of the near- to mid-infrared 
spectral slopes of $\rho$ Oph embedded sources: The solid line histogram represents sources with 
SED slopes, $a$, determined using mid-infrared data from this work; the dashed line histogram represents SED slopes determined from the 14.3 $\mu$m  {\it ISOCAM} photometry. The 2.2 $\mu$m photometry of Barsony et al. (1997) was used in both cases.  Note the large population
of Flat Spectrum ($-$0.3 $\le a \le +$0.3) objects.}
\label{Figure 4}
\end{figure}
\clearpage
\begin{figure}
\plotone{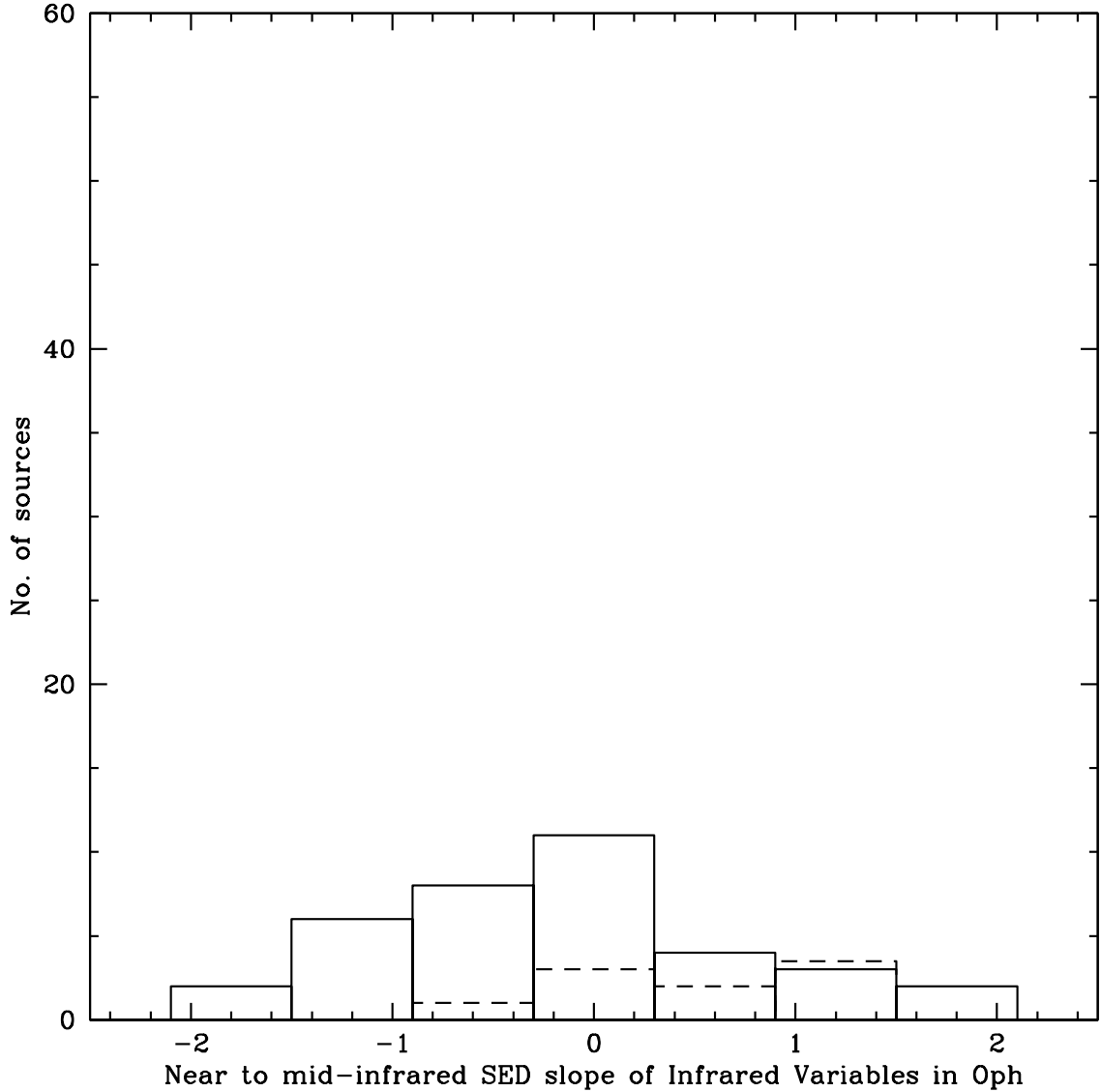}
\figcaption[f5.ps]{Histogram of the {\it observed} distribution of the near- to mid-infrared 
spectral slopes of $\rho$ Oph variables: The solid line represents this distribution for the mid-infrared variables
of Table 3; the dashed line represents the SED distribution of the near-infrared variables from Table 5 
of Barsony et al. (1997).  Note the tendency of the NIR variables towards earlier SED classes, whereas the mid-IR
variables seem to be evenly distributed through the SED classes with optically thick disks.}
\label{Figure 5}
\end{figure}
\clearpage
\begin{figure}
\plotone{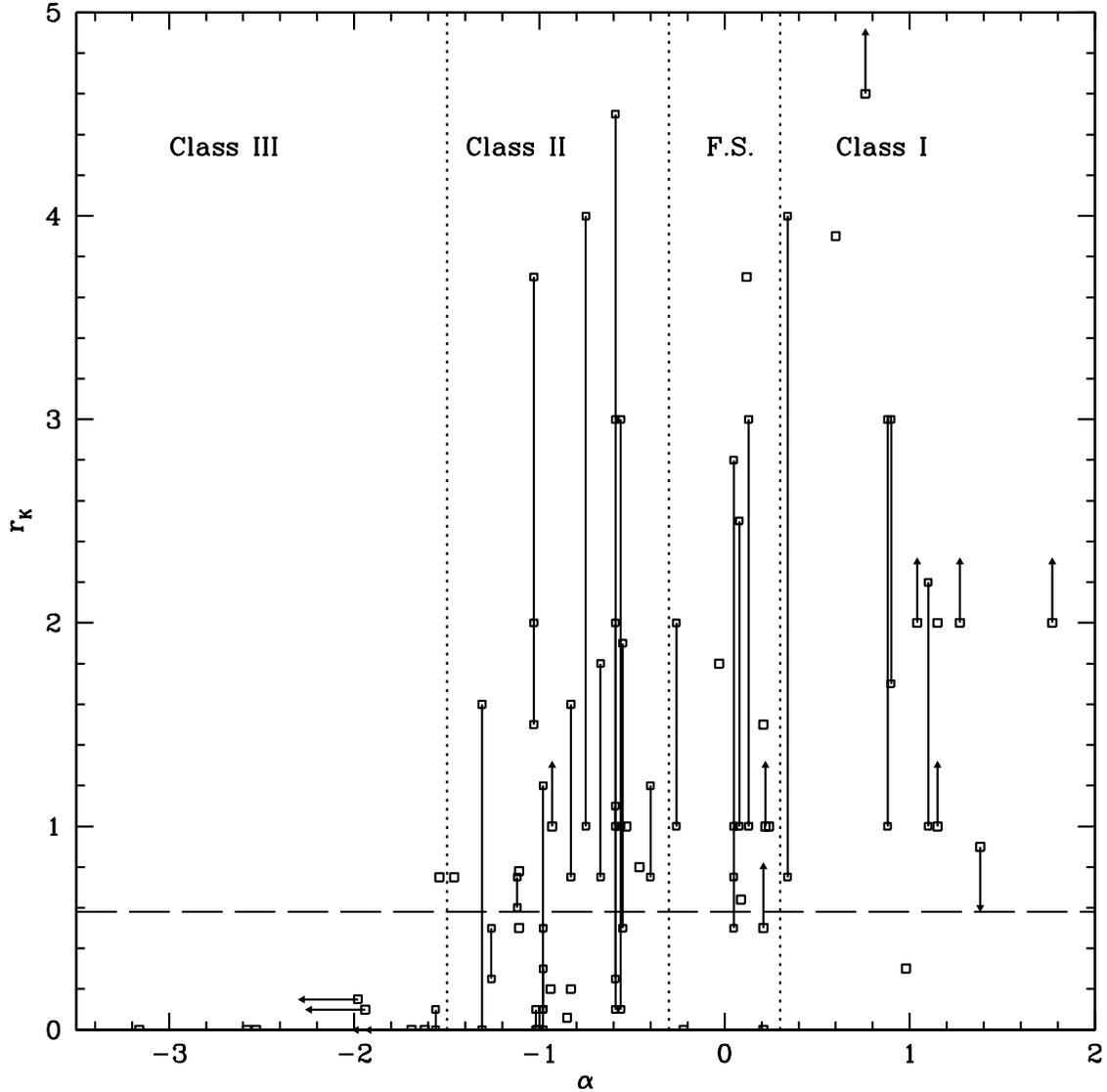}
\figcaption[f6.ps]{Plot of the K-band veiling (r$_K$ values) vs. the SED slopes, $\alpha$, for $\rho$ Oph embedded sources from Table 4.
Note that large variability in K-band veiling is a ubiquitous phenomenon amongst all SED classes with optically thick disks.
This is a robust result,  withstanding any errors that may have been introduced by the use of  non-simultaneous near- and mid-IR photometry in the determination of 
 the plotted SED slopes. The horizontal dashed line, drawn at r$_K=0.58$ indicates the dividing line between optically thick (above the line) and optically thin
(below the line) disks (Wilking et al. 2001). }
\label{Figure 6}
\end{figure}
\clearpage
\begin{figure}
\plotone{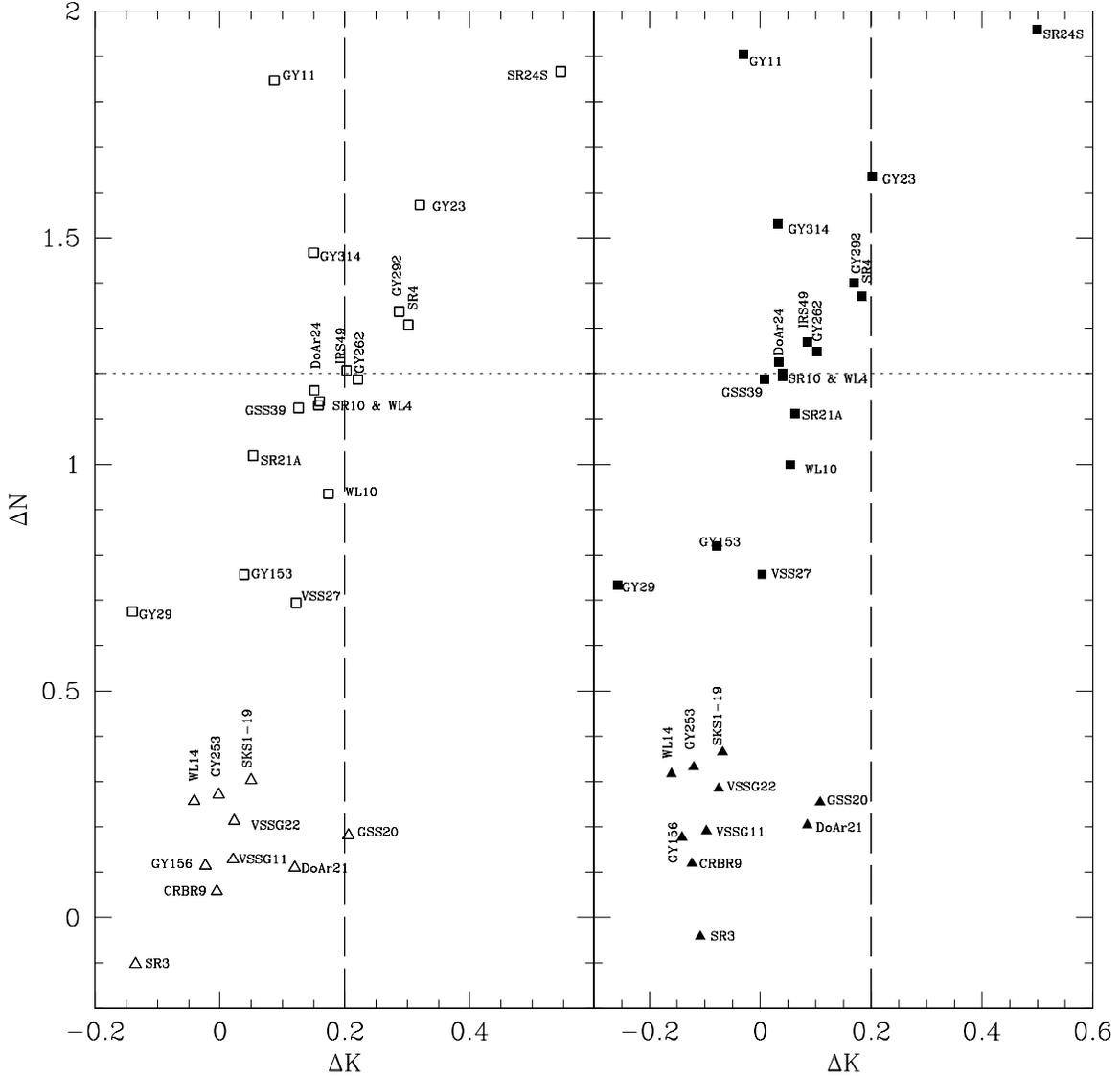}
\figcaption[f7.ps]{Plot of excess emission (above photospheric) at near-infrared ($\Delta K$)
and mid-infrared ($\Delta N$) wavelengths for Class II objects (squares) and Class III objects (triangles)
for which spectroscopic T$_{eff}$ measurements are available. The vertical dashed line at $\Delta K=0.2$ indicates the boundary between optically thick ($\Delta K \ge 0.2$) and optically thin
($\Delta K \le 0.2$) inner disks.  The horizontal, dotted line at $\Delta N =1.2$ indicates the boundary
between optically thick ($\Delta N \ge 1.2$) and optically thin ($\Delta N \le 1.2$) disk regions 
further out.  In the left panel, with open symbols, blackbody emission was used to model the central YSO,
whereas in the right panel, with filled symbols, Kurucz model atmospheres of the appropriate
effective temperature were used to model the central YSO.}
\label{Figure 7}
\end{figure}
\end{document}